\documentclass{webofc}
\usepackage[varg]{txfonts}   
\def\basf{\ensuremath{\texttt{basf2}}}

\begin{document}
\title{User documentation and training at Belle II\footnote{This work is licensed under a Creative Commons Attribution 4.0 International License (CC-BY 4.0).
The Creative Commons Attribution license allows re-distribution and re-use of a licensed work on the condition that the creator is appropriately credited. To view a copy of this license, visit \url{http://creativecommons.org/licenses/by/4.0/} or send a letter to Creative Commons, PO Box 1866, Mountain View, CA 94042, USA.

The terms of the Creative Commons license only apply to original material.
Reuse of material from other sources (marked within the Source) such as Diagrams, images, photos and text extracts may require further usage permits from the respective rights holder.}}
%
%

\author{
        \firstname{Sam} \lastname{Cunliffe}\inst{1}
        \and
        \firstname{Ilya} \lastname{Komarov}\inst{1}\fnsep\thanks{\email{ilya.komarov@desy.de}} \and
        \firstname{Thomas} \lastname{Kuhr}\inst{2} \and
        \firstname{Martin} \lastname{Ritter}\inst{2, 3} \and        
        \firstname{Francesco} \lastname{Tenchini}\inst{1}
        \ for the Belle II documentation and training group
}

\institute{The Deutsches Elektronen-Synchrotron 
\and
            Ludwig-Maximilians-Universität München
\and
            Origins Cluster
          }

\abstract{%
  Belle II is a rapidly growing collaboration with members from
one hundred and nineteen institutes spread around the globe. The software development team of
the experiment, as well as the software users, are very much
decentralised. Together with the active development of the software,
such decentralisation makes the adoption of the latest software
releases by users an essential, but quite challenging task.
To ensure the relevance of the documentation, we adopted the
policy of in-code documentation and configured a website that allows us to tie the documentation to given releases. To prevent tutorials from becoming
outdated, we covered them by unit-tests. For the user support, we use
a question and answer service that not only reduces repetition
of the same questions but also turned out to be a place for discussions
among the experts. A prototype of a metasearch engine for the different
sources of documentation has been developed. For training of the new users, we organise centralised StarterKit workshops attached to the collaboration meetings. The materials of the workshops are later used for self-education and organisation of local training sessions.
}
\maketitle
\section{Introduction}
\label{intro}
While the Belle II experiment is an ancestor of the Belle experiment, its analysis software framework (\basf) is completely new~\cite{basf2}. The framework consists of a \texttt{C++17} core with a \texttt{python3} user interface. 

In Summer 2017, the Belle II collaboration formed a \emph{Documentation, Training, and Software Outreach} working group whose main goals were to maintain analysis documentation, organise software tutorials and advertise the new tools. During the work on these goals, it soon became apparent that these tasks are highly correlated, and to achieve the goals we need to use a systematic approach that would cope with them simultaneously. With this, we reformulated the initial tasks themselves, so now they are to provide a smooth start for the newcomers and reliable support for the existing users.

\section{Belle II environment}
Two main aspects define the strategy of documentation and training efforts in Belle II: the speed of the software development and the structure of the collaboration. 

The software framework of the Belle II experiment is in the active development phase: there are $\sim1000$ commits per month with two major releases per year that can break backward compatibility. This forces users to migrate their analysis in order to be compatible with the newer software releases, and that should be communicated to the users. The analysis documentation should be quickly updated to simplify the migration.

As of February 2020, the Belle II collaboration has 1027 members from 119 institutes from 26 countries. The organisation is rapidly growing: several institutes and hundreds of new members join every year, which creates a demand for training. It is essential to account for the diverse language and cultural backgrounds of the collaboration members during the organisation of training sessions. The geographical spread of the collaboration implies difficulties in private communications with experts, so the official documentation becomes crucial for productive cooperation.

\section{Belle II training}
The training of the new members is done twice a year during three-day-long workshops called \emph{Belle II StarterKit Workshop} named after a similar idea in LHCb~\cite{LHCb}. The workshops are attached to the collaboration meetings that allows having a critical mass of experts and students. Each workshop is typically attended by 30 students and mentors and teachers body consists of around 20 experts. The workshops aim to teach participants the basics of the analysis in the Belle II collaboration. Throughout the workshop, students learn the Belle II-specific and the industry-standard software, some experimental physics and data analysis, and get an experience of the communication within the collaboration.

The workshop consists of lectures and hands-on sessions. Apart from the introduction to the software framework, the topics of lectures vary from event to event. Variation of the topics allows the collaboration to collect the asset of materials for self-education. When possible, we invite the lecturers from the young researchers who aim to become an expert in the covered topic since the preparation of the material is useful for the education of the lecturers. The lecturers rehearse in front of the organising committee and other lecturers a week before the workshop. Rehearsals allow to adjust the level and add cross-references to the talks.

The lectures devoted to the software are accompanied by Jupyter~\cite{Jupyter} notebooks. The notebooks are stored in a dedicated repository that is also accessible through a Jupyterhub server provided by DESY. The current Jupyter notebooks offer tutorials on basic python, the Belle II software framework, and basics of the data analysis.

In the very beginning of the workshop, all participants are split into working groups, consisting of the mentor and one or two students. The mentors are young researchers. Many of them are alumni of the previous workshops. The mentor takes care of the technical issues during the workshop, works as the first point of contact for the students, follows the progress of students through the lectures and organises discussions of the covered material, and coordinates the work of the group during the hands-on sessions. Such formation allows creating personal relations between the mentor and the students. It facilitates communication and can help to overcome language and cultural barriers. Moreover, many students find it more comfortable to ask questions in a more private environment than ask questions in front of the full room.

During the first hands-on session, students and mentors choose the task for the workshop. Typically, this is a simple analysis that involves work with analysis software framework and some further data analysis. Mentors can pick such exercises from several prepared ones. It can also happen that students have already started to work on the analysis. In this case, the mentor helps the student to work on it during the workshop.

After each workshop, we collect feedback from organisers and students. The students evaluate the accessibility and relevance of the material through the anonymous survey. The organisers share their experience during the face-to-face meeting and discuss the results of the students' survey. Such a feedback loop allows to improve the workshop and adjust the material to the current needs of the collaboration.

\section{Belle II documentation}
\subsection{Using Sphinx}
Fast growth and rapid changes in the experiment's software pushed us to search for alternatives for the traditional stand-alone documentation. In the period from 2017 to 2019, we moved all of the documentation of the user-facing code inside the software itself. The documentation obtained from in-code comments and dedicated text files is then rendered as an HTML page with the Sphinx~\cite{Sphinx} tool and users access it through the web.

\begin{figure}[h]
\centering
\includegraphics[width=0.49\linewidth, clip]{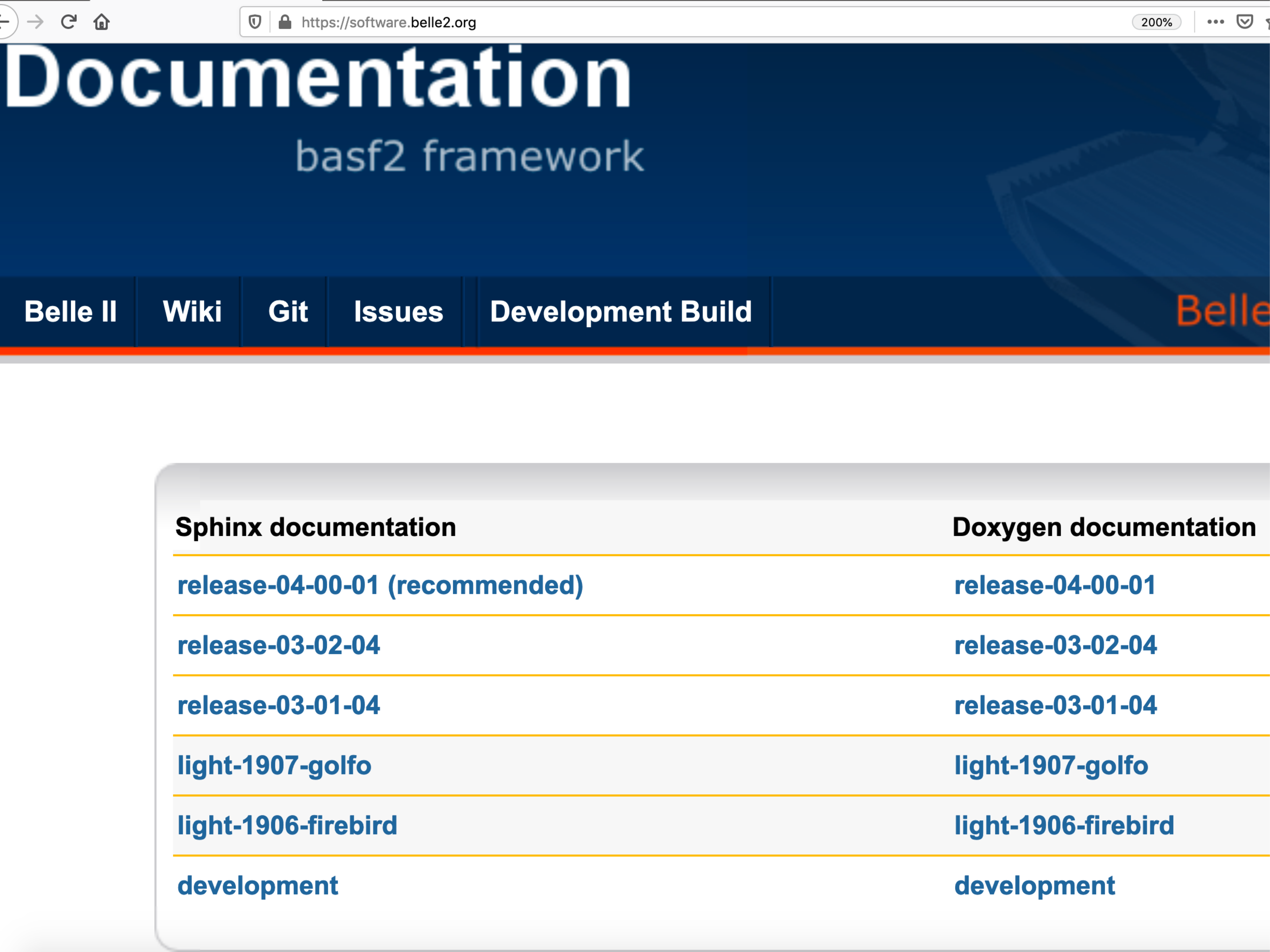}
\includegraphics[width=0.49\linewidth, clip]{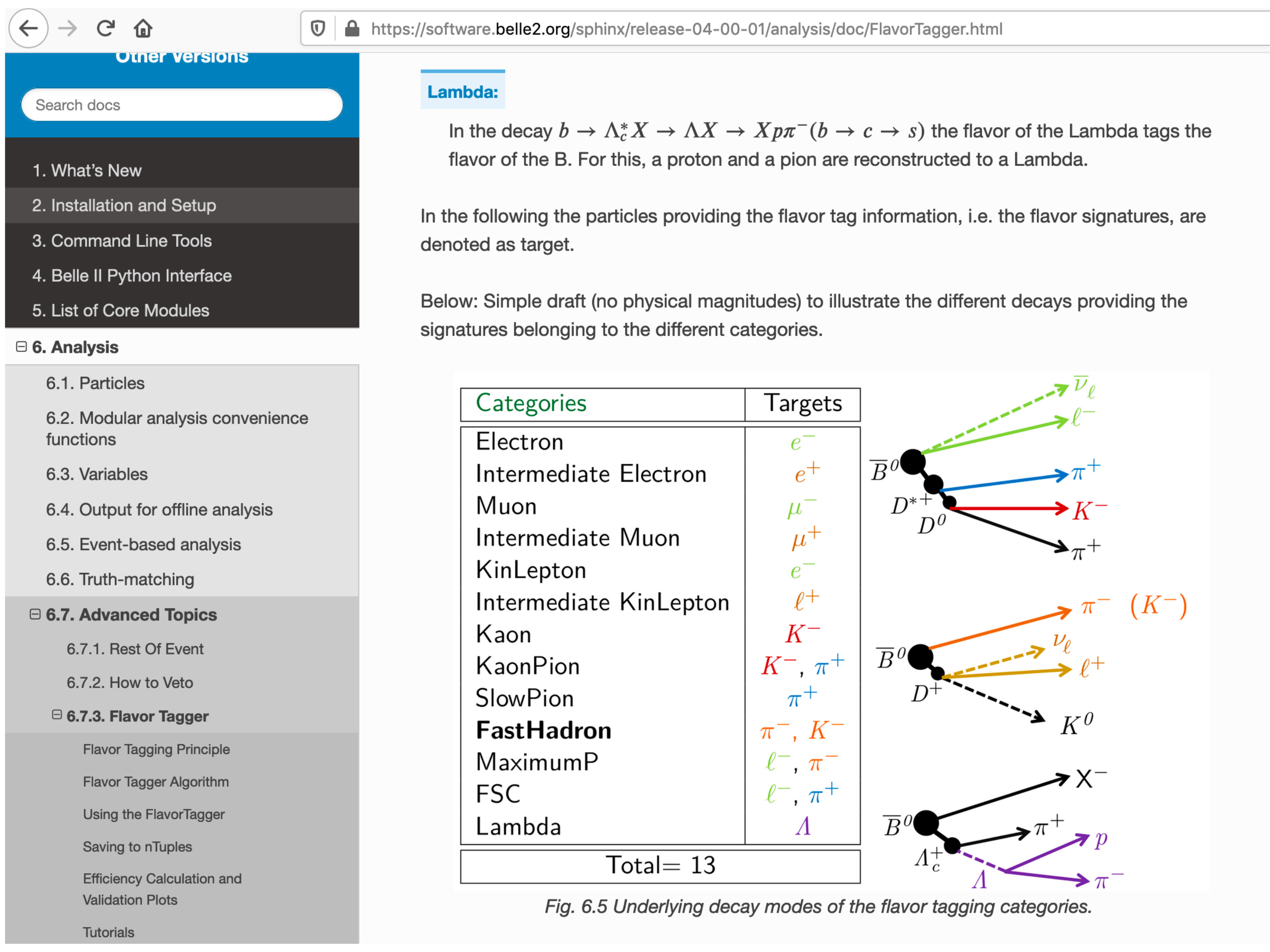}
\caption{Landing page of the documentation server with choise of the release (left) and screenshot of the user documentation renderred by Sphinx (right) }
\label{sphinx}       
\end{figure}

Using information from the comments for the user documentation prevents the outdating of the documentation. Moreover, such an approach allows supporting the documentation of the several releases simultaneously: the Sphinx server keeps the documentation of several releases, and the user can pick the relevant one. The user interface of the documentation server is shown in the Figure~\ref{sphinx}. 

\subsection{Code examples}
Another important part of user documentation is code examples. The code examples are the small collection of analysis scripts that demonstrates the main features of the analysis software framework. These examples are part of the software package and can become outdated. Unlike the in-code comments, the examples are isolated in the separate directory, and both the author of the changes and the reviewers might be unaware that introduced changes affect the examples. To cope with that issue, we covered the software examples with unit tests. As a result of this, if a specific pull request introduces changes that break any of the examples, this pull request cannot be merged. 

\begin{figure}[h]
\centering
\includegraphics[width=\linewidth, trim={3cm 10cm 0 0},clip]{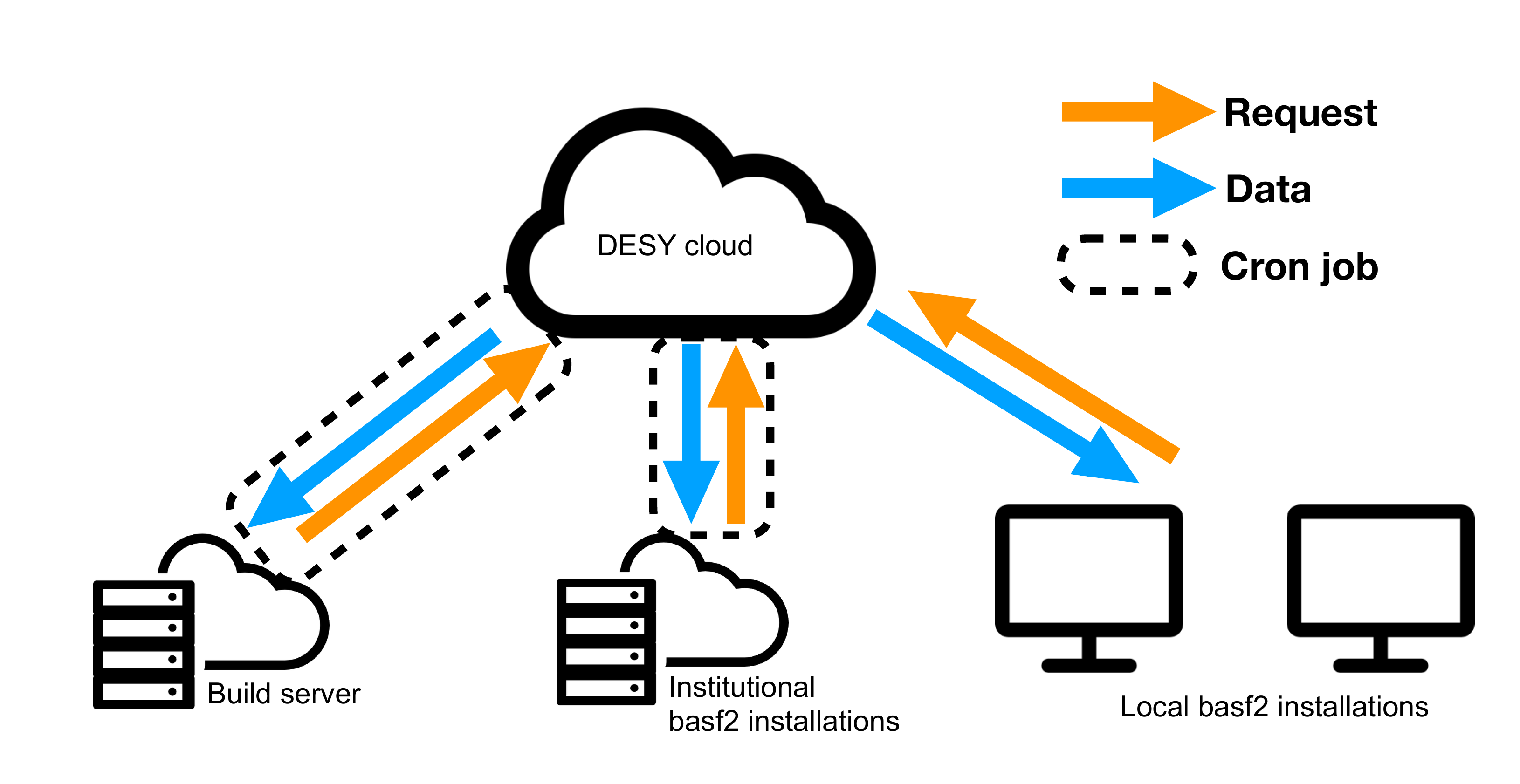}
\caption{Scheme of the distribution of the example datasets. For the build server and the institutional installations, the request of the new datasets is done through the cron jobs, while it is done manually for the local installations.}
\label{cloud}       
\end{figure}

The user examples run on the example datasets that cannot be included in the software repository due to their size. To ensure the reproducibility of the examples on different machines (local user installations, institutional servers, build server), we implemented a cloud-based solution using Nextcloud~\cite{Nextcloud} instance, as shown in the Figure~\ref{cloud}. We store the example datasets on the cloud server maintained by DESY as a part of institutional commitments.  It is the duty of the system administrators of the servers with instances of \basf to run cron jobs that synchronise them with local datasets. Such an approach allows updating the example files once they get updated centrally.

\subsection{User support}
\begin{figure}[h]
\centering
\includegraphics[width=\linewidth, clip]{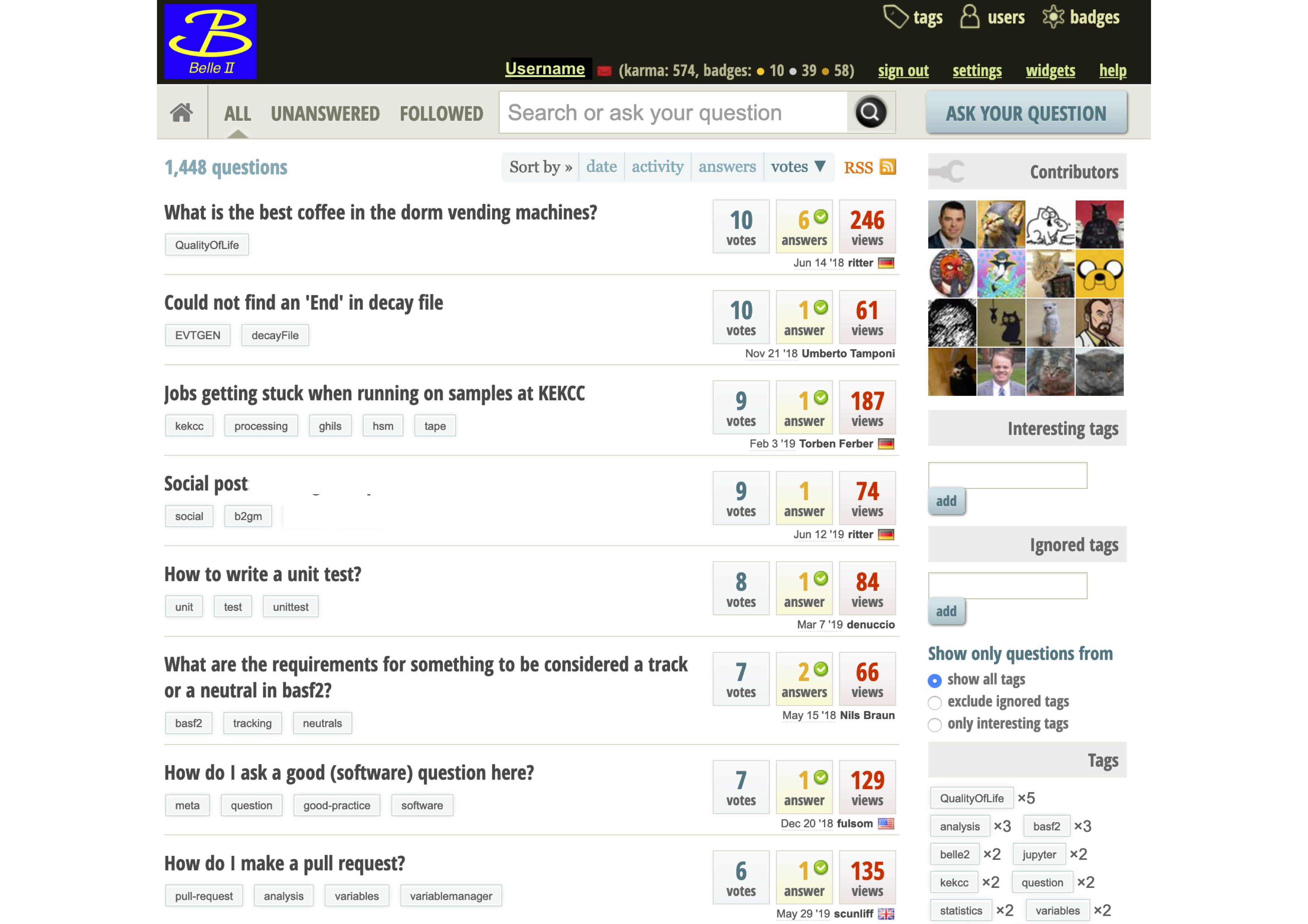}
\caption{Screenshot of the askbot instance.}
\label{askbot}       
\end{figure}

We use an Askbot~\cite{Askbot} instance to provide user support. Askbot turned out to be an excellent alternative to the traditional web archives of mailing lists. In 15 months, we accumulated 1233 questions, of which $83\%$ are answered with a typical time faster than one day. Askbot is used by 644 users of whom 240 either posed or answered the question. Askbot also offers a friendly user interface, as shown in the Figure~\ref{askbot} and the search engine.

The information required by Belle II collaborators is stored in code (and showed at the sphinx server), at a question and answer forum, and at a collaboration wiki that is mostly used for organisational purposes. A prototype of a metasearch engine for these three sources has been developed. The engine is used by the web page that works as a single entry point for the whole documentation requests.

\section{Conclusion}
The Belle II collaboration has made a concerted effort to improving the user documentation and providing dedicated, targeted, training workshops. The core of the training efforts is the Belle II StarterKit Workshops. They provide not only the essential knowledge and skills to start physics analysis in the Belle II collaboration but also a collection of training materials that can be used for self-education or for reproducing the events locally.

We solved the problem of documentation coverage and relevance by moving to in-code documentation that is later rendered to user-friendly format using the Sphinx tool. We use a cloud-based solution to ensure the validity of example, datasets among the different instances of \basf installations.


\end{document}